\documentclass[11pt]{article}

\usepackage[final]{acl}

\usepackage{times}
\usepackage{latexsym}
\usepackage[T1]{fontenc}
\usepackage[utf8]{inputenc}
\usepackage{microtype}
\usepackage{inconsolata}
\usepackage{graphicx}

\usepackage{booktabs}
\usepackage{bm}
\usepackage{multirow}
\usepackage{adjustbox}
\usepackage{enumitem}
\usepackage[most]{tcolorbox}
\usepackage{xcolor}
\usepackage{colortbl}
\usepackage{amsmath}
\usepackage{fancyvrb}

\usepackage{titlesec}
\titlespacing*{\section}{0pt}{6pt}{2pt}
\titlespacing*{\subsection}{0pt}{4pt}{2pt}
\titlespacing*{\subsubsection}{0pt}{3pt}{1pt}
\usepackage{caption}
\usepackage{tikz}
\usepackage{pgfplots}
\usetikzlibrary{patterns,positioning,shapes,arrows.meta,calc,er}
\usepackage[noend]{algpseudocode}
\usepackage{algorithmicx,algorithm}

\definecolor{darkblue}{rgb}{0.0, 0.0, 0.55}
\definecolor{lightgray}{rgb}{0.929, 0.929, 0.929}

\title{HIERA: Hierarchical Multi-Agent Relevance Assessment for Content Discovery Systems}

 \author{
   \textbf{Pritom Saha Akash}, \textbf{Phanideep Gampa}, \textbf{Chao Shen},
   \textbf{Ying Chen}, \textbf{Sheikh Muhammad Sarwar}
  \\
   Amazon
  \\
   \texttt{\{sapritom,phanide,shencha,yingchm,smsarwar\}@amazon.com}
  }

\begin{document}
\maketitle

\begin{abstract}
Content discovery systems depend on relevance judgment for search quality evaluation, but human annotation faces inter-annotator disagreement and scaling costs. While Large Language Models show promise as automated assessors, current approaches rely on flat aggregation strategies: single-step prompting, voting ensembles, or uncoordinated multi-agent pipelines that aggregate independent outputs without integration. We propose HIERA, a hierarchical multi-agent relevance assessment framework with four specialized agents: a Relevance Judge, Query Analyzer, Item Analyzer, and Relation Analyzer. The Judge determines when specialist analysis is needed; the Relation Analyzer then coordinates query and item analyses with external knowledge to establish relevance relationships for final judgment. Ablation studies show that the same agents and external knowledge without hierarchical coordination degrade performance, confirming that the coordination structure itself accounts for the improvement. Evaluation across five datasets (EVS, MSRD, ESCI, WANDS, Home Depot) shows improvements over 11 baselines: 10.2\% on Home Depot, 4.8\% on ESCI, and up to 38\% on EVS ($p < 0.05$). Hierarchical coordination yields 12.7\% improvement over uncoordinated collaboration using identical agents.
\end{abstract}

\section{Introduction}
\label{sec:introduction}

Content discovery systems help users navigate vast repositories across streaming platforms and e-commerce marketplaces. Their effectiveness depends on \textbf{relevance assessment}: determining whether content matches user intent. These assessments underpin performance metrics (NDCG, MAP) \cite{jarvelin2002cumulated,schutze2008introduction} and training data for recommendation models \cite{sanderson2010test,faggioli2023perspectives}, directly impacting user satisfaction and platform success.

However, obtaining reliable assessments faces two challenges. First, relevance is inherently subjective, with annotators frequently disagreeing on labels \cite{voorhees1998variations,bailey2008relevance}. Second, annotating thousands of query-item pairs across domains is economically infeasible \cite{thomas2024large}, forcing systems to operate with limited ground truth \cite{reddy2022shopping,deldjoo2020recommender}.


\begin{figure}[t]
    \centering
    \includegraphics[width=0.8\columnwidth]{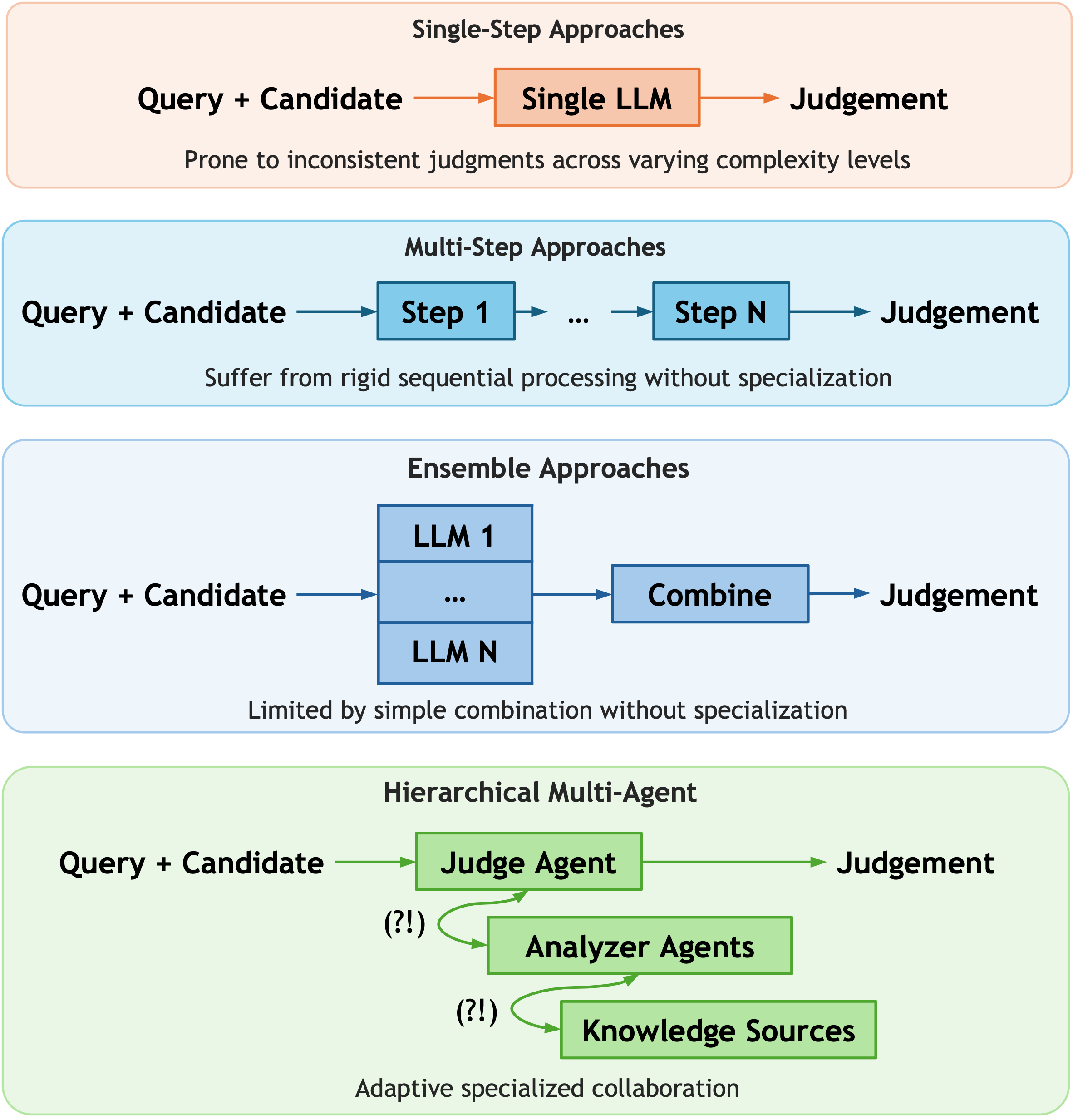}
    \caption{HIERA vs. other LLM-based approaches.}
    \label{fig:framework-comparison}
\end{figure}

To address these challenges, Large Language Models (LLMs) have emerged as promising alternatives for automated relevance 
assessment \cite{faggioli2023perspectives,thomas2024large}. These models can generate assessments at scale with reasonable human agreement \cite{soviero2024chatgpt,thomas2024large}. As shown in Figure \ref{fig:framework-comparison}, existing LLM-based approaches fall into three main paradigms. \textbf{Single-step approaches} use direct prompting to evaluate relevance in one step \cite{thomas2024large,soviero2024chatgpt,mehrdad2024large}, but compress all reasoning into a single inference, limiting their ability to decompose complex relevance relationships. \textbf{Multi-step approaches} employ sequential processing such as binary filtering followed by fine-grained classification \cite{schnabel2025multi}, or progressive refinement through multiple passes \cite{hosseini2025retrieve}, yet process information in isolation across stages without integrating insights between components. \textbf{Ensemble approaches} combine outputs from multiple models through voting or 
blending \cite{rahmani2025judgeblender,qian2025enhancing}, but lack coordination between models and specialized analysis of different relevance dimensions.

Despite these advances, existing LLM approaches face notable limitations when applied to content discovery systems. Traditional information retrieval matches explicit queries to documents based on topical relevance \cite{sanderson2010test}, while content discovery requires interpreting implicit user intent across multi-faceted items \cite{tsagkias2021challenges}. Consider the query \textit{``gray dresser''} matched against \textit{``ajinkya 7 drawer dresser''} with \textit{``metallic mercury finish.''} The ground truth labels this as relevant, yet all existing approaches consistently classify it as only partially relevant. This systematic underestimation occurs because the case requires simultaneously interpreting color relationships (metallic mercury as gray), assessing style compatibility (ornate versus simple), and evaluating functional matching (storage needs). Single-step approaches compress these into one inference and miss the connections, multi-step pipelines eliminate the item during binary filtering before domain interpretation can occur, and ensemble methods combine multiple underestimations since no individual model captures the cross-dimensional relationships.

This points to three requirements unmet by current approaches. First, \textbf{specialized analysis}: interpreting color relationships, style compatibility, and functional matching each require distinct analytical expertise. Second, \textbf{structured coordination}: these analyses must inform each other, as color interpretation depends on item characterization, which depends on query intent. Third, \textbf{evidence integration}: final judgments must synthesize findings from multiple perspectives with external knowledge, rather than aggregating independent assessments.

To address these limitations, we propose \textbf{HIERA} (HIErarchical Relevance Assessment), a multi-agent framework that decomposes relevance assessment into specialized, coordinating components. As illustrated in Figure \ref{fig:framework-comparison}, HIERA uses a \textbf{three-level hierarchy} where a \textit{Relevance Judge} determines when specialist analysis is needed and delegates to a \textit{Relation Analyzer}, which consults a \textit{Query Analyzer} for intent interpretation and an \textit{Item Analyzer} for candidate characterization, then synthesizes their outputs with external knowledge into a structured relevance argument returned to the Judge for final assessment.

Our contributions are threefold. First, we introduce HIERA, a hierarchical multi-agent framework for relevance assessment that enables specialized analysis, structured coordination, and evidence integration. Second, we demonstrate empirically that this hierarchical coordination outperforms flat aggregation (voting, blending, uncoordinated multi-agent) across five datasets (ESCI, WANDS, Home Depot, MSRD, and EVS), with improvements of 38\% on EVS and 10.2\% on Home Depot over the strongest baselines. Third, we demonstrate that providing agents with additional information or additional perspectives degrades performance without appropriate coordination structure, establishing that coordination topology determines whether scaling agents and knowledge sources helps or hurts.

\section{Related Work}
\label{sec:related_work}

\subsection{LLM-based Relevance Judgment}

\noindent\textbf{Single-Step Direct Assessment.} Early LLM-based relevance judgment employed direct prompting where models evaluate query-document relevance in a single inference step. Faggioli et al. \cite{faggioli2023perspectives} established the theoretical foundation, with subsequent work demonstrating human-comparable accuracy \cite{thomas2024large}, consistent performance at scale \cite{upadhyay2024large}, and successful application to e-commerce \cite{soviero2024chatgpt,mehrdad2024large}. However, single-step approaches suffer from \textit{reasoning compression}: they collapse all analytical reasoning into a single inference, preventing the decomposition needed to evaluate distinct aspects of complex relevance relationships.

\noindent\textbf{Multi-Stage Sequential Processing.} To address single-step limitations, researchers developed multi-stage pipelines that decompose relevance assessment into sequential components. Schnabel et al. \cite{schnabel2025multi} introduced binary filtering followed by fine-grained classification, showing improved accuracy over monolithic approaches. Farzi and Dietz \cite{farzi2024best,farzi2025criteria} demonstrated how criteria-based evaluation stages enhance judgment accuracy, while Hosseini et al. \cite{hosseini2025retrieve} developed multimodal frameworks combining textual and visual analysis stages. These approaches achieve better performance through \textit{analytical decomposition}, but suffer from \textit{stage isolation}: stages either operate independently or pass forward filtered outputs without the analytical reasoning behind them, limiting the integration needed when relevance depends on connecting query intent, item properties, and contextual knowledge.

\noindent\textbf{Ensemble and Collaborative Methods.} Recent work has explored collaborative approaches to overcome single-model limitations. Rahmani et al. \cite{rahmani2025judgeblender} developed JudgeBlender, demonstrating that combining multiple LLM judges through ensemble methods consistently outperforms individual models by mitigating biases and improving robustness. Sachdev et al. \cite{sachdev2024automated} integrated retrieval-augmented generation for content discovery contexts. These collaborative methods address individual model limitations through \textit{perspective diversity}, but suffer from \textit{expertise uniformity}: they combine general-purpose models without specialized domain knowledge, and simple aggregation methods cannot synthesize specialized insights needed for complex relevance relationships.

These approaches lack mechanisms for coordinating specialized expertise, motivating hierarchical architectures that organize agents by analytical function with structured information flow.

\subsection{LLM Agents}

Large Language Models have been enhanced with agent capabilities for complex reasoning through environmental interaction. Early single-agent systems (AutoGPT \cite{autogpt2023}, ReAct \cite{yao2023react}, QueryAgent \cite{huang2024queryagent}) demonstrated autonomous problem-solving through tool access but lack collaborative mechanisms for problems requiring diverse expertise.

Multi-agent frameworks have emerged where specialized agents collaborate on complex tasks. MetaGPT \cite{hong2023metagpt} simulates software company structures with role-based collaboration, AutoGen \cite{wu2024autogen} provides agent conversation orchestration, and AgentVerse \cite{chen2023agentverse} explores emergent behaviors in multi-agent systems. In evaluation contexts, Agent-as-a-Judge \cite{zhuge2024agent} employs agents for assessment tasks, while CollabEval \cite{qian2025enhancing} demonstrates how multi-agent collaboration enhances evaluation quality through peer-based consensus. However, these approaches target general reasoning rather than content discovery's specific challenges. No existing approach provides analytical specialization tailored to relevance assessment: distinct expertise for query interpretation, item characterization, and relationship analysis coordinated through hierarchical synthesis.

\section{Task Definition}
\label{sec:task_definition}

The \textit{relevance judgment} task involves assessing the degree of relevance between user queries and search results in content discovery systems such as \textbf{streaming services} and \textbf{e-commerce platforms}. Formally, given a query $q \in \mathcal{Q}$ expressing user information needs and a search result $c \in \mathcal{C}$ with associated metadata $\mathcal{M}(c)$, the task is to determine a relevance score $r \in \mathcal{R}$ indicating the degree of match between the query and result.

Query-result pairs in content discovery require multi-dimensional reasoning rather than uniform matching. Cases involving explicit attribute matching, such as ``\textit{action movies 2023}'' $\rightarrow$ ``\textit{Top Gun: Maverick}'', permit evaluation through direct criteria comparison. Cases requiring semantic interpretation, such as ``\textit{gray dresser}'' $\rightarrow$ ``\textit{metallic mercury finish dresser}'', demand reasoning across multiple dimensions: recognizing implicit relationships like color equivalence, assessing style compatibility, and verifying functional matching. This multi-dimensional nature poses challenges for current automated approaches, which typically lack mechanisms for decomposing the assessment into specialized analyses and integrating their findings, leading to suboptimal accuracy on cases requiring coordinated reasoning across query intent, item properties, and contextual knowledge.

\section{Methodology}
\label{sec:methodology}

\subsection{Framework Overview}

\begin{figure}[t]
    \centering
    \includegraphics[width=0.5\textwidth]{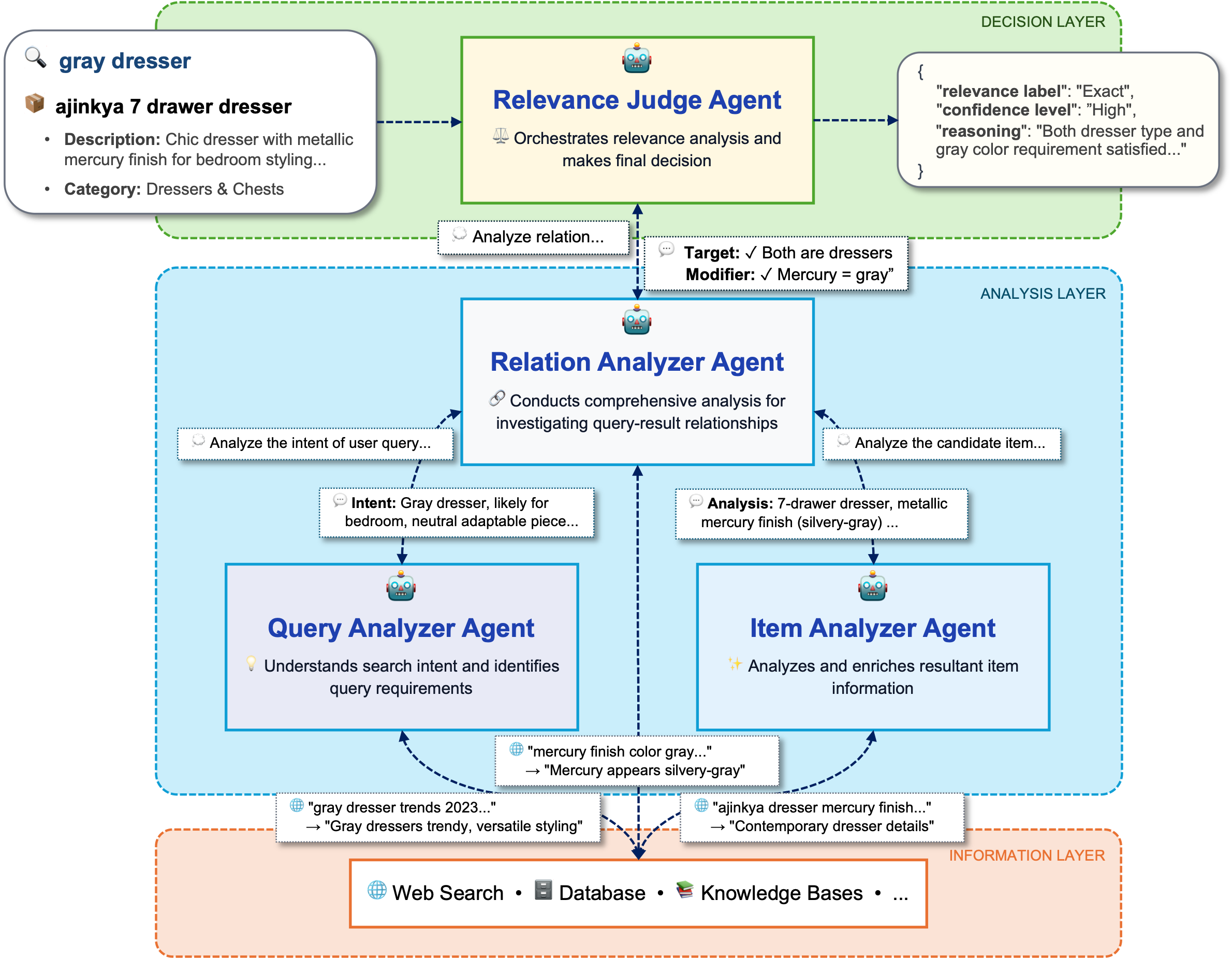}
    \caption{HIERA Framework}
    \label{fig:hiera_framework}
\end{figure}

Figure~\ref{fig:hiera_framework} illustrates HIERA, which decomposes relevance assessment into specialized analytical roles that integrate query understanding, item characterization, and relationship analysis through hierarchical coordination rather than independent aggregation.

The framework operates through a \textit{three-layer} architecture. The Decision Layer contains the \textit{Relevance Judge Agent} ($\mathcal{J}$), which determines what evidence is required and orchestrates consultation. The Analysis Layer provides three specialists: the \textit{Relation Analyzer} ($\mathcal{R}$) coordinates by consulting the \textit{Query Analyzer} ($\mathcal{Q}$) for query intent and the \textit{Item Analyzer} ($\mathcal{I}$) for candidate properties. The Information Layer enables external knowledge access ($\mathcal{K}$) for factual verification.

\noindent\textbf{Hierarchical Coordination Architecture.} Agents coordinate through a three-layer hierarchy where information flows between layers through tool-based consultation:
\begin{align}
\text{Decision Layer}: &\quad \mathcal{J} \rightarrow \{\mathcal{R}\} \\
\text{Analysis Layer}: &\quad [\mathcal{R} \leftrightarrow \{\mathcal{Q}, \mathcal{I}\}] \rightarrow \mathcal{K} \\
\text{Information Layer}: &\quad \mathcal{K}
\end{align}

This hierarchical design enables structured evidence synthesis: each specialist's output informs subsequent analytical steps, and the Relation Analyzer integrates findings into a coherent relevance argument before final judgment.

\subsection{Framework Components}

\subsubsection{Decision Layer}

Human experts decompose relevance assessment into distinct analytical steps: understanding user intent, characterizing candidate properties, and reasoning about their relationship. They consult specialized knowledge and synthesize findings before judgment, reflecting how structured coordination produces more reliable assessments than monolithic evaluation.

The \textit{Relevance Judge Agent} ($\mathcal{J}$) embodies this pattern. $\mathcal{J}$ examines available query-candidate evidence, assesses which analytical dimensions require specialist input, and formulates guidance for $\mathcal{R}$ accordingly. When all dimensions are clear, $\mathcal{J}$ produces a direct judgment.

\noindent\textbf{Multi-Dimensional Evidence Assessment.} $\mathcal{J}$ evaluates three dimensions when assessing whether specialist consultation is needed. \textit{Query Interpretability} evaluates whether user intent is explicitly stated or requires interpretation, as queries often contain implicit preferences beyond literal text. \textit{Information Completeness} reflects whether available candidate information provides sufficient evidence for confident judgment, as ambiguous product descriptions may demand clarification. \textit{Relational Transparency} determines whether query-result semantic connections are immediately apparent, as cases requiring reasoning about relationships such as color equivalence or material compatibility prompt external verification.

These dimensions do not imply one another: a query may be clear while item information is incomplete, or vice versa. $\mathcal{J}$ engages $\mathcal{R}$ whenever any dimension presents insufficient evidence. Figure~\ref{fig:hiera_framework} illustrates this with \textit{``gray dresser''} $\rightarrow$ \textit{``metallic mercury finish dresser''}: $\mathcal{J}$ recognizes clear query interpretability but limited relational transparency, prompting consultation with $\mathcal{R}$, which integrates query intent (from $\mathcal{Q}$), item characterization (from $\mathcal{I}$), and external evidence (from $\mathcal{K}$) into a coherent relevance argument before $\mathcal{J}$ produces the final judgment.

\subsubsection{Analysis Layer}

When the Decision Layer identifies that consultation is needed, resolving gaps requires distinct expertise: extracting explicit requirements and implicit intent from queries, identifying stated and inferred attributes from candidates, and mapping semantic connections across dimensions. This separation reflects the single responsibility principle: agents perform more effectively when focused on specific tasks rather than handling diverse analytical challenges simultaneously.

The Analysis Layer addresses this through three specialized agents: the \textit{Relation Analyzer Agent} ($\mathcal{R}$) performs connection analysis and coordinates specialist consultation, the \textit{Query Analyzer Agent} ($\mathcal{Q}$) extracts explicit requirements and implicit intent, and the \textit{Item Analyzer Agent} ($\mathcal{I}$) identifies both explicit and inferred candidate characteristics. The hierarchy is realized through $\mathcal{R}$, which coordinates specialist analyses and synthesizes their outputs into a structured relevance argument before returning evidence to $\mathcal{J}$.

\noindent\textbf{Relation Analyzer Agent ($\mathcal{R}$).} $\mathcal{R}$ analyzes the semantic connections between query and candidate while coordinating specialist consultation. Unlike peer-based multi-agent approaches that aggregate independent agent outputs, $\mathcal{R}$ acts as a hierarchical coordinator: it consults $\mathcal{Q}$ when query intent is uncertain, $\mathcal{I}$ when candidate information is insufficient, and $\mathcal{K}$ for factual verification, then synthesizes these findings into a structured relevance argument. For the ``gray dresser'' case, $\mathcal{R}$ identifies that the color relationship requires both query intent analysis and item characteristic verification: it consults $\mathcal{Q}$ to clarify the implicit preference for furniture, $\mathcal{I}$ to verify that ``metallic mercury finish'' exhibits gray-toned visual properties, and $\mathcal{K}$ to confirm the color equivalence in furniture contexts. The resulting argument is returned to $\mathcal{J}$ for final judgment.

\noindent\textbf{Query Analyzer Agent ($\mathcal{Q}$).} $\mathcal{Q}$ focuses on query understanding, activated when $\mathcal{R}$ identifies insufficient query interpretability. It decomposes queries into explicit requirements (stated constraints) and implicit intent (unstated preferences for style or functionality). When contextual understanding is needed, $\mathcal{Q}$ accesses $\mathcal{K}$ for clarification, returning findings to $\mathcal{R}$.

\noindent\textbf{Item Analyzer Agent ($\mathcal{I}$).} $\mathcal{I}$ focuses on candidate characterization, activated when $\mathcal{R}$ identifies insufficient candidate information. It extracts explicit characteristics (directly stated features) and inferred properties (implicit attributes such as color appearance or style category). When verification is needed, $\mathcal{I}$ accesses $\mathcal{K}$, returning findings to $\mathcal{R}$.

\noindent\textbf{Coordination Protocol.} When $\mathcal{J}$ determines that confident assessment requires additional evidence, it delegates to $\mathcal{R}$ with guidance on identified gaps. $\mathcal{R}$ performs its own connection analysis and, based on the gaps encountered, consults specialists: $\mathcal{Q}$ for query interpretation, $\mathcal{I}$ for candidate characterization, and $\mathcal{K}$ for factual verification. Critically, this is integrative rather than independent: $\mathcal{R}$ consults specialists based on identified gaps and synthesizes their findings into a single structured argument before returning it to $\mathcal{J}$.

This coordination pattern follows:
\begin{align}
\text{Gaps} &= \mathcal{R}(q, c, \text{guidance}_{\mathcal{J}}) \\
\text{Active} &= \{s \in \{\mathcal{Q}, \mathcal{I}, \mathcal{K}\} : s \text{ addresses Gaps}\} \\
\text{Result} &= \text{integrate}(\bigcup_{s \in \text{Active}} s(\text{Gaps}))
\end{align}

This hierarchical structure differs from peer-based multi-agent approaches where agents operate independently and outputs are aggregated post-hoc. Because all specialist findings pass through $\mathcal{R}$ before reaching $\mathcal{J}$, the final relevance argument reflects how query intent, item properties, and external evidence relate to each other rather than treating them as independent signals.

\subsubsection{Information Layer}

The Information Layer ($\mathcal{K}$) extends HIERA's analytical capabilities beyond parametric knowledge when Analysis Layer agents identify information gaps. $\mathcal{K}$ provides external knowledge access to support agent analysis when internal reasoning is insufficient for establishing relevance relationships.

\noindent\textbf{Knowledge Integration.} Analysis Layer agents ($\mathcal{R}$, $\mathcal{Q}$, $\mathcal{I}$) access $\mathcal{K}$ when they identify information gaps that cannot be resolved through internal reasoning. $\mathcal{K}$ provides factual grounding needed to establish semantic connections that require domain-specific verification. The layer supports diverse knowledge sources including structured databases and web search to accommodate domain-specific requirements.

\noindent\textbf{Structured Knowledge Access.} Agents formulate targeted queries to $\mathcal{K}$ based on specific gaps identified during their analysis. Retrieved information is returned to the requesting agent, which integrates it into its ongoing analytical process before passing findings up through the hierarchy. This design ensures that external knowledge is contextualized within the agent's analytical frame rather than appended as raw supplementary text.

\section{Experimental Setup}
\label{sec:experimental_setup}

\subsection{Datasets}

We evaluate HIERA across five content discovery datasets spanning \textit{entertainment search} (EVS, MSRD) and \textit{product search} (ESCI, WANDS, Home Depot). For public datasets, we construct evaluation subsets of approximately 1,000 query-item pairs through random sampling with fixed seed (42), ensuring equal representation of each relevance level.
\textbf{EVS} is a proprietary entertainment video search dataset (300 query-title pairs, 3-level scale) with rich structured metadata and domain-specific database access.
\textbf{MSRD}\footnote{https://github.com/metarank/msrd} contains binary relevance annotations for movie search covering actors, directors, genres, themes, and cultural references.
\textbf{ESCI} \cite{reddy2022shopping} provides e-commerce query-product pairs across diverse categories with 4-level taxonomy (Exact, Substitute, Complement, Irrelevant).
\textbf{WANDS} \cite{wands} covers furniture and home decor search with 3-level relevance (Exact Match, Partial Match, Irrelevant).
\textbf{Home Depot}\footnote{https://www.kaggle.com/c/home-depot-product-search-relevance/data} features home improvement product search with 3-level relevance across diverse query types including product names and technical specifications.

\subsection{Baselines}

We compare against 11 methods across three categories.
\textbf{Single-step:} (1) \textbf{Zero-Shot} \cite{faggioli2023perspectives} uses direct prompting without examples; (2) \textbf{Multi-Criteria} \cite{mehrdad2024large} evaluates multiple relevance dimensions through structured prompting; (3) \textbf{DNA-Prompt} \cite{thomas2024large} uses Descriptive-Narrative-Aspects structured reasoning.
\textbf{Multi-step:} (4) \textbf{Binary-Graded} \cite{rahmani2025judgeblender} applies binary filtering then fine-grained scoring; (5) \textbf{Multi-Stage} \cite{schnabel2025multi} refines judgments through progressive reasoning stages; (6) \textbf{Self-Instruct} \cite{soviero2024chatgpt} generates annotation guidelines from examples then applies them; (7) \textbf{RAG-MMR} \cite{sachdev2024automated} combines retrieval-augmented generation with diverse example selection; (8) \textbf{RAER} \cite{hosseini2025retrieve} generates query-specific evaluation criteria through iterative refinement.
\textbf{Ensemble/Collaboration:} (9) \textbf{Prompt-Blender} \cite{rahmani2025judgeblender} aggregates multiple prompt formulations; (10) \textbf{LLM-Blender} \cite{rahmani2025judgeblender} ensembles multiple LLMs with diverse strategies; (11) \textbf{CollabEval} \cite{qian2025enhancing} uses multi-agent collaborative evaluation with consensus building.

\subsection{Evaluation Metrics}

We evaluate all methods using three standard metrics for relevance assessment: Accuracy (\textbf{Acc}) measures the proportion of correctly classified query-result pairs, Macro F1-score (\textbf{F1}) computes F1-score for each relevance class independently and averages them ensuring balanced evaluation across relevance categories, and Cohen's $\kappa$ (\textbf{$\kappa$}) assesses agreement between predicted and ground truth labels while accounting for chance agreement. Statistical significance is determined using McNemar's Test for paired comparisons with Bonferroni correction ($p < 0.05$). Bootstrap 95\% confidence intervals for all ablation conditions are reported in Appendix~\ref{sec:confidence_intervals}.

\subsection{Implementation Details}

We implement HIERA using Claude 3.7 Sonnet ($\mathcal{J}$, $\mathcal{R}$) and Claude 3.5 Haiku ($\mathcal{Q}$, $\mathcal{I}$) via AWS Bedrock with temperature 0, using LangGraph\footnote{https://langchain-ai.github.io/langgraph/} with ReAct design pattern \cite{yao2023react}. All baselines use Claude 3.7 Sonnet for fair comparison; LLM-Blender and CollabEval use diverse model sets as required by their methodologies. Full implementation details including tool access configuration, baseline model assignments, and system prompts are provided in Appendix~\ref{sec:implementation_details}.

\section{Results}
\label{sec:results}
\subsection{Overall Performance}

Table~\ref{tab:entertainment_search_results} presents results for entertainment search datasets (EVS and MSRD), while Table~\ref{tab:product_search_results} shows product search performance (ESCI, WANDS, and Home Depot). All results include accuracy, F1-score, and Cohen's $\kappa$ metrics. Statistical significance is assessed using McNemar's test with Bonferroni correction (p < 0.05), with asterisks (*) indicating statistically significant differences compared to \textit{HIERA}. Note that Self-Instruct and RAG-MMR results are unavailable for EVS dataset due to the lack of few-shot samples required by these methods.

\begin{table}[htbp]
\caption{Entertainment results. \textbf{Bold}: best, *: $p{<}0.05$.}
\label{tab:entertainment_search_results}
\centering
\resizebox{\columnwidth}{!}{%
\begin{tabular}{l|ccc|ccc}
\toprule
& \multicolumn{3}{c|}{\textbf{EVS}} & \multicolumn{3}{c}{\textbf{MSRD}} \\
\textbf{Method} & \textbf{Acc} & \textbf{F1} & \textbf{$\kappa$} & \textbf{Acc} & \textbf{F1} & \textbf{$\kappa$} \\
\midrule
Zero-Shot & 0.517* & 0.586* & 0.271* & 0.762* & 0.753* & 0.524* \\
Multi-Criteria & 0.507* & 0.567* & 0.238* & 0.723* & 0.706* & 0.446* \\
DNA-Prompt & 0.557* & 0.606* & 0.292* & 0.739* & 0.726* & 0.478* \\
Binary-Graded & 0.400* & 0.468* & 0.136* & 0.810* & 0.540* & 0.620* \\
Multi-Stage & 0.427* & 0.361* & 0.156* & 0.768* & 0.517* & 0.550* \\
Self-Instruct & -- & -- & -- & 0.839 & 0.838* & 0.678* \\
RAG-MMR & -- & -- & -- & 0.821* & 0.819* & 0.642* \\
RAER & 0.533* & 0.594* & 0.286* & 0.768* & 0.759* & 0.536* \\
Prompt-Blender & 0.520* & 0.566* & 0.243* & 0.755* & 0.744* & 0.510* \\
LLM-Blender & 0.543* & 0.586* & 0.264* & 0.758* & 0.748* & 0.516* \\
CollabEval & 0.567* & 0.609* & 0.284* & 0.773* & 0.769* & 0.546* \\
\midrule
\rowcolor{blue!10} \textbf{HIERA (Ours)} & \textbf{0.713} & \textbf{0.699} & \textbf{0.461} & \textbf{0.859} & \textbf{0.858} & \textbf{0.718} \\
\bottomrule
\end{tabular}
}
\end{table}
\begin{table}[htbp]
\caption{Product search results. \textbf{Bold}: best; *: $p < 0.05$.}
\label{tab:product_search_results}
\centering
\resizebox{\columnwidth}{!}{%
\begin{tabular}{l|ccc|ccc|ccc}
\toprule
& \multicolumn{3}{c|}{\textbf{ESCI}} & \multicolumn{3}{c|}{\textbf{WANDS}} & \multicolumn{3}{c}{\textbf{Home Depot}} \\
\textbf{Method} & \textbf{Acc} & \textbf{F1} & \textbf{$\kappa$} & \textbf{Acc} & \textbf{F1} & \textbf{$\kappa$} & \textbf{Acc} & \textbf{F1} & \textbf{$\kappa$} \\
\midrule
Zero-Shot & 0.481* & 0.480* & 0.308* & 0.630* & 0.625* & 0.445* & 0.414* & 0.415* & 0.122* \\
Multi-Criteria & 0.401* & 0.405* & 0.201* & 0.654 & 0.647* & 0.481* & 0.421* & 0.415* & 0.132* \\
DNA-Prompt & 0.445* & 0.450* & 0.260* & 0.628* & 0.622* & 0.442* & 0.427* & 0.428* & 0.141* \\
Binary-Graded & 0.444* & 0.423* & 0.259* & 0.627* & 0.627* & 0.440* & 0.405* & 0.406* & 0.108* \\
Multi-Stage & 0.433* & 0.342* & 0.245* & 0.595* & 0.440* & 0.393* & 0.379* & 0.285* & 0.070* \\
Self-Instruct & 0.518 & 0.519* & 0.357* & 0.649* & 0.653* & 0.473* & 0.407* & 0.407* & 0.111* \\
RAG-MMR & 0.500* & 0.498* & 0.333* & 0.638* & 0.645* & 0.457* & 0.404* & 0.405* & 0.107* \\
RAER & 0.474* & 0.466* & 0.299* & 0.627* & 0.617* & 0.440* & 0.412* & 0.309* & 0.120* \\
Prompt-Blender & 0.472* & 0.475* & 0.296* & 0.648* & 0.644* & 0.472* & 0.414* & 0.415* & 0.122* \\
LLM-Blender & 0.428* & 0.430* & 0.237* & 0.662 & 0.663* & 0.493* & 0.416* & 0.418* & 0.125* \\
CollabEval & 0.480* & 0.482* & 0.307* & 0.630* & 0.638* & 0.445* & 0.410* & 0.399* & 0.116* \\
\midrule
\rowcolor{blue!10} \textbf{HIERA (Ours)} & \textbf{0.543} & \textbf{0.545} & \textbf{0.391} & \textbf{0.681} & \textbf{0.686} & \textbf{0.521} & \textbf{0.464} & \textbf{0.465} & \textbf{0.197} \\
\bottomrule
\end{tabular}
}
\end{table}

\textit{HIERA} demonstrates substantial and consistent improvements across all evaluation datasets, with particularly notable gains on challenging domains. The framework achieves its most dramatic improvement on the EVS dataset, outperforming the best baseline Zero-Shot by 38\% in accuracy (0.713 vs 0.517) and 70\% in Cohen's $\kappa$ (0.461 vs 0.271). On product search datasets, \textit{HIERA} consistently outperforms the strongest baselines: 4.8\% improvement over Self-Instruct on ESCI (0.543 vs 0.518), 4.3\% over LLM-Blender on WANDS (0.681 vs 0.654), and 10.2\% over Multi-Criteria on Home Depot (0.464 vs 0.421). Note that Home Depot exhibits low absolute agreement across all methods due to inherent annotation ambiguity in this dataset; the relative improvement remains substantial. Even on the relatively easier MSRD dataset, \textit{HIERA} achieves 2.4\% improvement over Self-Instruct (0.859 vs 0.839), demonstrating effectiveness across varying complexity levels.

Across baseline categories, single-step approaches show inconsistent performance (Zero-Shot: 0.517 on EVS vs 0.762 on MSRD), multi-step methods perform better with Self-Instruct as the strongest traditional baseline, and ensemble methods (LLM-Blender, CollabEval) show mixed results despite combining multiple perspectives. HIERA consistently outperforms all categories, indicating that structured coordination with specialized roles provides genuine advantages over both individual methods and aggregation-based approaches.

\subsection{Ablation Study}

To understand each component's contribution, we conduct ablation studies on MSRD and WANDS examining three aspects: (1) external knowledge integration through web search, (2) individual agent roles in the Analysis Layer, and (3) coordination through the Relation Analyzer. Table~\ref{tab:ablation_results} presents results across component combinations.

\begin{table}[htbp]
\centering
\caption{Ablation Study Results}
\label{tab:ablation_results}
\resizebox{1\columnwidth}{!}{%
\begin{tabular}{l|ccc|ccc}
\toprule
& \multicolumn{3}{c|}{\textbf{MSRD}} & \multicolumn{3}{c}{\textbf{WANDS}} \\
\textbf{Variant} & \textbf{Acc} & \textbf{F1} & \textbf{$\kappa$} & \textbf{Acc} & \textbf{F1} & \textbf{$\kappa$} \\
\midrule
Judge Only & 0.747 & 0.737 & 0.494 & 0.622 & 0.627 & 0.433 \\
Judge + Web Search & 0.752 & 0.743 & 0.504 & 0.617 & 0.620 & 0.425 \\
\midrule
Judge + Relation \\ Analyzer & 0.748 & 0.738 & 0.496 & 0.668 & 0.673 & 0.501 \\
Judge + Relation \\ Analyzer + Web Search & 0.778 & 0.774 & 0.556 & 0.655 & 0.661 & 0.482 \\
\midrule
Judge + Query\&Item \\ Analyzers & 0.741 & 0.730 & 0.482 & 0.627 & 0.627 & 0.440 \\
Judge + Query\&Item \\ Analyzers + Web Search & 0.732 & 0.719 & 0.463 & 0.621 & 0.623 & 0.431 \\
\midrule
Judge + All Analyzers & 0.762 & 0.755 & 0.524 & 0.659 & 0.664 & 0.488 \\
\rowcolor{green!10} \textbf{HIERA (Full System)} & \textbf{0.859} & \textbf{0.858} & \textbf{0.718} & \textbf{0.681} & \textbf{0.686} & \textbf{0.521} \\
\bottomrule
\end{tabular}%
}
\end{table}

The ablation shows three patterns. First, external knowledge alone provides minimal benefits: adding web search to Judge-only yields negligible change (0.747 to 0.752 on MSRD, 0.622 to 0.617 on WANDS). Second, the Relation Analyzer is the crucial component: its inclusion yields substantial improvements (0.748 on MSRD, 0.668 on WANDS), amplified further with web search (0.778, 0.655), while Q and I without R show limited effectiveness and even degradation with web search (0.741 to 0.732 on MSRD). Third, the full system (0.859 on MSRD, 0.681 on WANDS) substantially outperforms the complete agent ensemble without coordination (0.762, 0.659). The Judge already performs structured multi-dimensional reasoning within a 3,000-token budget, and in the Judge + All Analyzers condition it additionally receives specialist analyses as context, comparable to extended chain-of-thought with multiple analytical perspectives. The persistent gap confirms that the improvement requires interactive synthesis through $\mathcal{R}$: both configurations provide the same agents and knowledge, but only the full system routes consultation through $\mathcal{R}$ for synthesis before returning evidence to $\mathcal{J}$.

\section{Analysis and Discussion}
\label{sec:analysis}
Having established \textit{HIERA}'s superior performance through quantitative evaluation and ablation studies, we now provide deeper analysis across three key dimensions: (1) the coordination mechanisms that enable multi-agent collaboration, (2) the effectiveness of external knowledge integration strategies, and (3) computational cost analysis.

\subsection{Multi-Agent Coordination Patterns}

The Relation Analyzer serves as the central synthesizer: it receives guidance from $\mathcal{J}$, consults specialists as needed, and integrates their findings into a structured relevance argument. This contrasts with ensemble methods where agents operate independently and outputs are aggregated post-hoc. Three behavioral patterns characterize this coordination:
(1) \textbf{Role-based specialization}: each agent addresses a specific analytical dimension ($\mathcal{Q}$ for query intent, $\mathcal{I}$ for candidate properties, $\mathcal{R}$ for relationships), operating within a focused scope rather than reasoning about all dimensions simultaneously.
(2) \textbf{Hierarchical synthesis}: $\mathcal{R}$ synthesizes specialist findings into a coherent argument that accounts for interactions between query intent, item properties, and external evidence, distinguishing the full system from the uncoordinated ablation where agents produce independent analyses without integration.
(3) \textbf{Knowledge integration}: external knowledge improves performance only when integrated through the coordination hierarchy; as shown in Table~\ref{tab:websearch_concat}, the same web search results that improve HIERA (+12.7\% on MSRD) degrade most baselines when concatenated directly.

\noindent\textbf{Activation frequency.} $\mathcal{J}$ delegates to the full hierarchy in 91.5\% of cases, with only 8.5\% receiving direct judgment. This is expected: the system is designed as a coordination architecture, not an adaptive router. The contribution is not selective activation but the coordination protocol itself, as confirmed by the ablation: the same cases processed by all agents without hierarchical coordination (Judge + All Analyzers) achieve 37\% lower $\kappa$ despite identical coverage. The value lies in how specialists are coordinated through $\mathcal{R}$, not in when they are activated.

\begin{table}[h]
\centering
\caption{External knowledge integration effectiveness.}
\label{tab:websearch_concat}
\resizebox{1\columnwidth}{!}{%
\begin{tabular}{l|ccc|ccc}
\toprule
& \multicolumn{3}{c|}{\textbf{MSRD}} & \multicolumn{3}{c}{\textbf{WANDS}} \\
\textbf{Method} & \textbf{Acc} & \textbf{F1} & \textbf{$\kappa$} & \textbf{Acc} & \textbf{F1} & \textbf{$\kappa$} \\
\midrule
Zero-Shot & \textcolor{blue!60!black}{+0.3\%} & \textcolor{blue!60!black}{+0.4\%} & \textcolor{blue!60!black}{+0.8\%} & \textcolor{blue!60!black}{+1.4\%} & \textcolor{blue!60!black}{+1.8\%} & \textcolor{blue!60!black}{+3.0\%} \\
Multi-Criteria & \textcolor{red!60!black}{-2.4\%} & \textcolor{red!60!black}{-2.9\%} & \textcolor{red!60!black}{-7.6\%} & \textcolor{red!60!black}{-5.6\%} & \textcolor{red!60!black}{-5.6\%} & \textcolor{red!60!black}{-11.0\%} \\
DNA-Prompt & \textcolor{red!60!black}{-0.4\%} & \textcolor{red!60!black}{-0.5\%} & \textcolor{red!60!black}{-1.3\%} & \textcolor{red!60!black}{-1.3\%} & \textcolor{red!60!black}{-1.3\%} & \textcolor{red!60!black}{-2.7\%} \\
Binary-Graded & \textcolor{red!60!black}{-0.5\%} & \textcolor{red!60!black}{-0.5\%} & \textcolor{red!60!black}{-1.4\%} & \textcolor{red!60!black}{-1.6\%} & \textcolor{red!60!black}{-1.6\%} & \textcolor{red!60!black}{-3.4\%} \\
Multi-Stage & \textcolor{red!60!black}{-5.6\%} & \textcolor{red!60!black}{-8.0\%} & \textcolor{red!60!black}{-18.2\%} & \textcolor{blue!60!black}{+1.2\%} & \textcolor{blue!60!black}{+0.9\%} & \textcolor{blue!60!black}{+2.6\%} \\
Self-Instruct & \textcolor{red!60!black}{-0.7\%} & \textcolor{red!60!black}{-0.6\%} & \textcolor{red!60!black}{-1.8\%} & \textcolor{gray!70!black}{+0.0\%} & \textcolor{blue!60!black}{+0.2\%} & \textcolor{gray!70!black}{+0.0\%} \\
RAG-MMR & \textcolor{red!60!black}{-5.4\%} & \textcolor{red!60!black}{-5.8\%} & \textcolor{red!60!black}{-13.7\%} & \textcolor{blue!60!black}{+3.0\%} & \textcolor{blue!60!black}{+2.0\%} & \textcolor{blue!60!black}{+6.2\%} \\
RAER & \textcolor{red!60!black}{-1.3\%} & \textcolor{red!60!black}{-1.3\%} & \textcolor{red!60!black}{-3.7\%} & \textcolor{blue!60!black}{+0.6\%} & \textcolor{blue!60!black}{+1.7\%} & \textcolor{blue!60!black}{+1.4\%} \\
Prompt-Blender & \textcolor{blue!60!black}{+0.4\%} & \textcolor{blue!60!black}{+0.5\%} & \textcolor{blue!60!black}{+1.2\%} & \textcolor{red!60!black}{-0.2\%} & \textcolor{red!60!black}{-0.1\%} & \textcolor{red!60!black}{-0.3\%} \\
LLM-Blender & \textcolor{blue!60!black}{+0.8\%} & \textcolor{blue!60!black}{+0.8\%} & \textcolor{blue!60!black}{+2.3\%} & \textcolor{red!60!black}{-1.5\%} & \textcolor{red!60!black}{-2.0\%} & \textcolor{red!60!black}{-3.0\%} \\
CollabEval & \textcolor{blue!60!black}{+1.2\%} & \textcolor{blue!60!black}{+0.9\%} & \textcolor{blue!60!black}{+3.3\%} & \textcolor{red!60!black}{-6.5\%} & \textcolor{red!60!black}{-7.0\%} & \textcolor{red!60!black}{-13.8\%} \\
\midrule
\rowcolor{blue!10} \textbf{HIERA} & \textbf{\textcolor{blue!50!black}{+12.7\%}} & \textbf{\textcolor{blue!50!black}{+13.6\%}} & \textbf{\textcolor{blue!50!black}{+37.0\%}} & \textbf{\textcolor{blue!50!black}{+3.3\%}} & \textbf{\textcolor{blue!50!black}{+3.3\%}} & \textbf{\textcolor{blue!50!black}{+6.8\%}} \\
\bottomrule
\end{tabular}%
}
\end{table}
\subsection{External Knowledge Integration Impact}

To test whether coordination provides genuine advantages over simpler knowledge augmentation, we provide all baseline methods with identical web search results used by HIERA, concatenated directly to their input prompts (Table~\ref{tab:websearch_concat}). This controls for information availability, isolating the effect of integration strategy. Most methods degrade despite accessing the same information: Multi-Criteria (-2.4\% on MSRD, -5.6\% on WANDS), Multi-Stage (-5.6\% on MSRD), and even CollabEval shows inconsistent behavior (+1.2\% on MSRD, -6.5\% on WANDS). In contrast, HIERA achieves consistent improvements (+12.7\% on MSRD, +3.3\% on WANDS), demonstrating that effective knowledge integration requires coordination rather than concatenation.

\subsection{Computational Cost Analysis}

\noindent\textbf{Token Usage Analysis.} We analyze token consumption on MSRD to assess whether gains arise from coordination structure rather than increased computation. HIERA consumes 2,917 tokens per case at the orchestration level. The uncoordinated ablation (Judge + All Analyzers) consumes 2,866 tokens, within 1.8\% of HIERA's budget, yet scores 37\% lower in Cohen's $\kappa$ (0.524 vs.\ 0.718). The Judge alone achieves $\kappa$ = 0.494 within a 3,000-token budget; enriching it with all specialist analyses yields only $\kappa$ = 0.524, confirming that coordination topology, not token volume or richer context, accounts for the gain. For comparison, CollabEval broadcasts full evaluation histories across multiple discussion rounds, consuming 9,621 tokens while achieving lower agreement ($\kappa$ = 0.546).

\noindent\textbf{Latency.} HIERA achieves mean response times of 35.6s on MSRD and 35.9s on WANDS, compared to 17.7--20.2s for CollabEval and 0.93--0.98s for single-step methods. However, HIERA remains 50$\times$ faster than human experts (35s vs 30 minutes per judgment \cite{sachdev2024automated}), making it practical for offline evaluation tasks where accuracy requirements justify the computational investment.

Qualitative case study analysis (Appendix~\ref{sec:case_studies}) reveals that HIERA excels at detecting constraint violations through coordinated specialist consultation (e.g., language mismatches, functional misalignments), but can amplify errors when agents reinforce initial misinterpretations in cases requiring subjective cultural judgment.

\section{Conclusion}
\label{sec:conclusion}

We present \textit{HIERA}, a hierarchical multi-agent framework for relevance assessment in content discovery systems. Evaluation across five datasets demonstrates consistent improvements over 11 baselines: 38\% on EVS, 10.2\% on Home Depot, and statistically significant gains across all datasets. Ablation studies confirm that the coordination structure itself accounts for the improvement: the same agents and knowledge access without hierarchical coordination score 37\% lower in Cohen's $\kappa$, and external knowledge that improves performance under coordination degrades most baselines when provided directly. These findings demonstrate that structured orchestration through a central synthesizer outperforms both independent multi-agent operation and flat aggregation strategies for relevance assessment tasks requiring multi-dimensional reasoning. For practitioners building multi-agent evaluation systems, the implication is clear: investing in coordination protocols yields larger gains than adding more agents or more expensive models.

\section*{Limitations}
We acknowledge several limitations of this work. First, one of our evaluation datasets (EVS) is proprietary, limiting full reproducibility; however, we evaluate on four public datasets to enable independent verification. Second, HIERA incurs higher latency than single-step approaches (35.6s vs under 1s), though it remains 50$\times$ faster than human annotation and operates within the same order of magnitude as other multi-agent approaches such as CollabEval (37.1s). Third, our evaluation focuses on English-language content discovery; generalization to multilingual settings remains unexplored. Finally, the Judge's consultation decisions rely on LLM reasoning, which may exhibit variability across model versions or repeated runs; we mitigate this with temperature 0 but do not formally evaluate decision stability across multiple runs. Additionally, all experiments use Claude models; generalization to other LLM families (e.g., GPT-4, Gemini) remains to be validated.

\bibliography{references}

@inproceedings{faggioli2023perspectives,
  title={Perspectives on large language models for relevance judgment},
  author={Faggioli, Guglielmo and Dietz, Laura and Clarke, Charles LA and Demartini, Gianluca and Hagen, Matthias and Hauff, Claudia and Kando, Noriko and Kanoulas, Evangelos and Potthast, Martin and Stein, Benno and others},
  booktitle={Proceedings of the 2023 ACM SIGIR International Conference on Theory of Information Retrieval},
  pages={39--50},
  year={2023}
}

@article{upadhyay2024large,
  title={A Large-Scale Study of Relevance Assessments with Large Language Models: An Initial Look},
  author={Upadhyay, Shivani and Pradeep, Ronak and Thakur, Nandan and Campos, Daniel and Craswell, Nick and Soboroff, Ian and Dang, Hoa Trang and Lin, Jimmy},
  journal={arXiv preprint arXiv:2411.08275},
  year={2024}
}

@inproceedings{thomas2024large,
  title={Large language models can accurately predict searcher preferences},
  author={Thomas, Paul and Spielman, Seth and Craswell, Nick and Mitra, Bhaskar},
  booktitle={Proceedings of the 47th International ACM SIGIR Conference on Research and Development in Information Retrieval},
  pages={1930--1940},
  year={2024}
}

@inproceedings{rahmani2025judgeblender,
  title={JudgeBlender: Ensembling Automatic Relevance Judgments},
  author={Rahmani, Hossein A and Yilmaz, Emine and Craswell, Nick and Mitra, Bhaskar},
  booktitle={Companion Proceedings of the ACM on Web Conference 2025},
  pages={1268--1272},
  year={2025}
}

@inproceedings{schnabel2025multi,
  title={Multi-stage large language model pipelines can outperform gpt-4o in relevance assessment},
  author={Schnabel, Julian A and Trippas, Johanne R and Scholer, Falk and Hettiachchi, Danula},
  booktitle={Companion Proceedings of the ACM on Web Conference 2025},
  pages={1288--1292},
  year={2025}
}

@inproceedings{soviero2024chatgpt,
  title={ChatGPT goes shopping: LLMs can predict relevance in ecommerce search},
  author={Soviero, Beatriz and Kuhn, Daniel and Salle, Alexandre and Moreira, Viviane Pereira},
  booktitle={European Conference on Information Retrieval},
  pages={3--11},
  year={2024},
  organization={Springer}
}

@article{mehrdad2024large,
  title={Large language models for relevance judgment in product search},
  author={Mehrdad, Navid and Mohapatra, Hrushikesh and Bagdouri, Mossaab and Chandran, Prijith and Magnani, Alessandro and Cai, Xunfan and Puthenputhussery, Ajit and Yadav, Sachin and Lee, Tony and Zhai, ChengXiang and others},
  journal={arXiv preprint arXiv:2406.00247},
  year={2024}
}

@inproceedings{sachdev2024automated,
  title={Automated Query-Product Relevance Labeling using Large Language Models for E-commerce Search},
  author={Sachdev, Jayant and D Rosario, Sean and Phatak, Abhijeet and Wen, He and Kirti, Swati and Tripathy, Chittaranjan},
  booktitle={Proceedings of the 2024 8th International Conference on Natural Language Processing and Information Retrieval},
  pages={32--40},
  year={2024}
}

@inproceedings{hosseini2025retrieve,
  title={Retrieve, Annotate, Evaluate, Repeat: Leveraging Multimodal LLMs for Large-Scale Product Retrieval Evaluation},
  author={Hosseini, Kasra and Kober, Thomas and Krapac, Josip and Vollgraf, Roland and Cheng, Weiwei and Peleteiro Ramallo, Ana},
  booktitle={European Conference on Information Retrieval},
  pages={149--163},
  year={2025},
  organization={Springer}
}

@article{qian2025enhancing,
  title={Enhancing LLM-as-a-judge via multi-agent collaboration},
  author={Qian, Yiyue and Zhang, Shinan and Zhou, Yun and Ding, Haibo and Socolinsky, Diego and Zhang, Yi},
  year={2025}
}

@article{reddy2022shopping,
title={Shopping Queries Dataset: A Large-Scale {ESCI} Benchmark for Improving Product Search},
author={Chandan K. Reddy and Lluís Màrquez and Fran Valero and Nikhil Rao and Hugo Zaragoza and Sambaran Bandyopadhyay and Arnab Biswas and Anlu Xing and Karthik Subbian},
year={2022},
eprint={2206.06588},
archivePrefix={arXiv}
}

@InProceedings{wands,  
  title = {WANDS: Dataset for Product Search Relevance Assessment},  
  author = {Chen, Yan and Liu, Shujian and Liu, Zheng and Sun, Weiyi and Baltrunas, Linas and Schroeder, Benjamin},  
  booktitle = {Proceedings of the 44th European Conference on Information Retrieval},  
  year = {2022},  
  numpages = {12}  
}

@article{jarvelin2002cumulated,
  title={Cumulated gain-based evaluation of IR techniques},
  author={J{\"a}rvelin, Kalervo and Kek{\"a}l{\"a}inen, Jaana},
  journal={ACM Transactions on Information Systems (TOIS)},
  volume={20},
  number={4},
  pages={422--446},
  year={2002},
  publisher={ACM New York, NY, USA}
}

@book{schutze2008introduction,
  title={Introduction to information retrieval},
  author={Sch{\"u}tze, Hinrich and Manning, Christopher D and Raghavan, Prabhakar},
  volume={39},
  year={2008},
  publisher={Cambridge University Press Cambridge}
}

@article{sanderson2010test,
  title={Test collection based evaluation of information retrieval systems},
  author={Sanderson, Mark and others},
  journal={Foundations and Trends{\textregistered} in Information Retrieval},
  volume={4},
  number={4},
  pages={247--375},
  year={2010},
  publisher={Now Publishers, Inc.}
}

@inproceedings{voorhees1998variations,
  title={Variations in relevance judgments and the measurement of retrieval effectiveness},
  author={Voorhees, Ellen M},
  booktitle={Proceedings of the 21st annual international ACM SIGIR conference on Research and development in information retrieval},
  pages={315--323},
  year={1998}
}

@inproceedings{bailey2008relevance,
  title={Relevance assessment: are judges exchangeable and does it matter},
  author={Bailey, Peter and Craswell, Nick and Soboroff, Ian and Thomas, Paul and de Vries, Arjen P and Yilmaz, Emine},
  booktitle={Proceedings of the 31st annual international ACM SIGIR conference on Research and development in information retrieval},
  pages={667--674},
  year={2008}
}

@article{deldjoo2020recommender,
  title={Recommender systems leveraging multimedia content},
  author={Deldjoo, Yashar and Schedl, Markus and Cremonesi, Paolo and Pasi, Gabriella},
  journal={ACM Computing Surveys (CSUR)},
  volume={53},
  number={5},
  pages={1--38},
  year={2020},
  publisher={ACM New York, NY, USA}
}

@inproceedings{tsagkias2021challenges,
  title={Challenges and research opportunities in ecommerce search and recommendations},
  author={Tsagkias, Manos and King, Tracy Holloway and Kallumadi, Surya and Murdock, Vanessa and De Rijke, Maarten},
  booktitle={ACM Sigir Forum},
  volume={54},
  number={1},
  pages={1--23},
  year={2021},
  organization={ACM New York, NY, USA}
}

@article{farzi2024best,
  title={Best in tau@ llmjudge: Criteria-based relevance evaluation with llama3},
  author={Farzi, Naghmeh and Dietz, Laura},
  journal={arXiv preprint arXiv:2410.14044},
  year={2024}
}

@inproceedings{farzi2025criteria,
  title={Criteria-Based LLM Relevance Judgments},
  author={Farzi, Naghmeh and Dietz, Laura},
  booktitle={Proceedings of the 2025 International ACM SIGIR Conference on Innovative Concepts and Theories in Information Retrieval (ICTIR)},
  pages={254--263},
  year={2025}
}

@article{zhuge2024agent,
  title={Agent-as-a-judge: Evaluate agents with agents},
  author={Zhuge, Mingchen and Zhao, Changsheng and Ashley, Dylan and Wang, Wenyi and Khizbullin, Dmitrii and Xiong, Yunyang and Liu, Zechun and Chang, Ernie and Krishnamoorthi, Raghuraman and Tian, Yuandong and others},
  journal={arXiv preprint arXiv:2410.10934},
  year={2024}
}

@inproceedings{yao2023react,
  title={React: Synergizing reasoning and acting in language models},
  author={Yao, Shunyu and Zhao, Jeffrey and Yu, Dian and Du, Nan and Shafran, Izhak and Narasimhan, Karthik and Cao, Yuan},
  booktitle={International Conference on Learning Representations (ICLR)},
  year={2023}
}

@inproceedings{wu2024autogen,
  title={Autogen: Enabling next-gen LLM applications via multi-agent conversations},
  author={Wu, Qingyun and Bansal, Gagan and Zhang, Jieyu and Wu, Yiran and Li, Beibin and Zhu, Erkang and Jiang, Li and Zhang, Xiaoyun and Zhang, Shaokun and Liu, Jiale and others},
  booktitle={First Conference on Language Modeling},
  year={2024}
}

@article{chen2023agentverse,
  title={Agentverse: Facilitating multi-agent collaboration and exploring emergent behaviors in agents},
  author={Chen, Weize and Su, Yusheng and Zuo, Jingwei and Yang, Cheng and Yuan, Chenfei and Qian, Chen and Chan, Chi-Min and Qin, Yujia and Lu, Yaxi and Xie, Ruobing and others},
  journal={arXiv preprint arXiv:2308.10848},
  volume={2},
  number={4},
  pages={6},
  year={2023}
}

@misc{autogpt2023,
  title={AutoGPT: Build, Deploy, and Run AI Agents},
  author={AutoGPT-Team},
  year={2023},
  url={https://github.com/Significant-Gravitas/Auto-GPT},
  note={GitHub repository}
}

@inproceedings{hong2023metagpt,
  title={MetaGPT: Meta programming for a multi-agent collaborative framework},
  author={Hong, Sirui and Zhuge, Mingchen and Chen, Jonathan and Zheng, Xiawu and Cheng, Yuheng and Wang, Jinlin and Zhang, Ceyao and Wang, Zili and Yau, Steven Ka Shing and Lin, Zijuan and others},
  booktitle={The Twelfth International Conference on Learning Representations},
  year={2023}
}

@article{huang2024queryagent,
  title={Queryagent: A reliable and efficient reasoning framework with environmental feedback-based self-correction},
  author={Huang, Xiang and Cheng, Sitao and Huang, Shanshan and Shen, Jiayu and Xu, Yong and Zhang, Chaoyun and Qu, Yuzhong},
  journal={arXiv preprint arXiv:2403.11886},
  year={2024}
}

\appendix
\section{Implementation Details}
\label{sec:implementation_details}

\subsection{HIERA Configuration}
The Decision Layer ($\mathcal{J}$) employs Claude 3.7 Sonnet for relevance judgment with 3000 max tokens, while the Analysis Layer uses Claude 3.5 Haiku for Query Analyzer ($\mathcal{Q}$) and Item Analyzer ($\mathcal{I}$) with 1000 max tokens each. Relation Analyzer ($\mathcal{R}$) uses Claude 3.7 Sonnet with 1500 max tokens for coordination tasks. All agents use temperature 0 for deterministic outputs.

\textbf{Framework Implementation:} HIERA is implemented using LangGraph with ReAct design pattern, enabling agents to interleave reasoning and tool usage in iterative cycles. Each agent follows: (1) Reasoning - analyze current state and determine next action, (2) Acting - execute tools or consult specialists, (3) Observing - process results and update state. LangGraph manages agent coordination through directed graphs where nodes represent agents and edges define communication pathways, with persistent state management across agent interactions.

\textbf{Tool Access:} Each agent has access to specific tools reflecting its role in the hierarchy. $\mathcal{J}$ can invoke $\mathcal{R}$ as a tool. $\mathcal{R}$ can invoke $\mathcal{Q}$, $\mathcal{I}$, and web search. $\mathcal{Q}$ and $\mathcal{I}$ can each invoke web search and database retrieval independently. For the EVS dataset, both web search and structured database access are available; for public datasets (MSRD, ESCI, WANDS, Home Depot), only web search is used. Web search returns 5 results per query. All experiments use consistent system prompts for each agent role, with fixed prompt templates maintained across all datasets.

\subsection{Baseline Configuration}
For baseline methods, most approaches use Claude 3.7 Sonnet with temperature 0.0 for fair comparison with HIERA's main decision-making agents, including Zero-Shot, Multi-Criteria, DNA-Prompt, Binary-Graded, Self-Instruct, RAG-MMR, RAER, and Prompt-Blender. The Multi-Stage baseline uses a two-model approach: Claude 3.5 Haiku for binary filtering (Stage 1) and Claude 3.7 Sonnet for fine-grained classification (Stage 2). LLM-Blender and CollabEval employ diverse model sets as required by their original methodologies: LLM-Blender uses three different models (Claude 3.7 Sonnet, Amazon Nova Pro, and Mistral Large) for ensemble diversity, while CollabEval uses four diverse models (Mistral Large, Claude 3.5 Haiku, Claude 3.5 Sonnet, and Amazon Nova Pro) for multi-agent evaluation with Claude 3.7 Sonnet as the final judge.

\section{Agent System Prompt Templates}
\label{sec:prompt_templates}

This document provides the complete system prompt templates used in HIERA's multi-agent architecture for full reproducibility. Each template serves as the system prompt for the respective agent, defining its role, capabilities, and behavioral constraints within the multi-agent coordination framework. All prompts utilize configurable parameters (shown in \textcolor{red!70!black}{red}) that are dynamically populated based on domain-specific configurations and dataset characteristics.

\subsection{Judge Agent System Prompt}
\label{subsec:judge_prompt}
The Judge Agent system prompt implements the three-dimensional evidence assessment and coordinates multi-agent consultation through the hierarchical framework.

\begin{figure}[t]
\begin{tcolorbox}[title=Judge Agent Prompt Template, colback=gray!3, colframe=gray!50!black, boxrule=0.5pt, left=4pt, right=4pt, top=2pt, bottom=2pt]
\scriptsize
\textbf{TASK:} Judge relevance between query and candidate result\\[2pt]
\textbf{DOMAIN:} \textcolor{red!70!black}{\{domain\}} \textcolor{red!70!black}{\{domain\_context\}}\\[2pt]
\textbf{JUDGMENT CRITERIA:} \textcolor{red!70!black}{\{judgment\_criteria\}}\\[3pt]
You are an expert relevance judge.\\[2pt]
Evaluate this case and determine your approach:\\
\hspace*{4pt}-- Assess whether you can make a confident judgment with available information\\
\hspace*{4pt}-- If straightforward, proceed with your assessment\\
\hspace*{4pt}-- If complex or uncertain, engage appropriate analytical support\\[2pt]
Your judgment guides the process:\\
\hspace*{4pt}-- Trust your expertise for clear cases\\
\hspace*{4pt}-- Seek specialist input when facing ambiguity\\
\hspace*{4pt}-- Adapt your thoroughness to match case complexity\\[3pt]
\textbf{AVAILABLE TOOLS:} \textcolor{red!70!black}{\{tool\_description\}}\\[3pt]
\textbf{OUTPUT FORMAT:} JSON only\\
\texttt{\{"relevance\_label": "\{labels\}",}\\
\texttt{~"confidence\_level": "high | medium | low",}\\
\texttt{~"reasoning": "explanation grounded on criteria"\}}\\[2pt]
\textbf{CRITICAL:} Respond with valid JSON only, no additional text.
\end{tcolorbox}
\caption{Judge Agent prompt template.}
\label{fig:judge_prompt}
\end{figure}

\subsection{Relation Analyzer Agent System Prompt}
\label{subsec:relation_analyzer_prompt}
The Relation Analyzer Agent system prompt defines the coordination specialist role, responsible for orchestrating specialist consultations and synthesizing relationship analysis.

\begin{figure}[t]
\begin{tcolorbox}[title=Relation Analyzer Agent Prompt Template, colback=gray!3, colframe=gray!50!black, boxrule=0.5pt, left=4pt, right=4pt, top=2pt, bottom=2pt]
\scriptsize
\textbf{TASK:} Analyze relationships between query and candidate result\\[2pt]
\textbf{DOMAIN:} \textcolor{red!70!black}{\{domain\}} \textcolor{red!70!black}{\{domain\_context\}}\\[2pt]
\textbf{JUDGMENT CONTEXT:} \textcolor{red!70!black}{\{judgment\_criteria\}}\\[3pt]
Provide comprehensive relationship analysis considering all available information and judgment criteria. Be concise and direct for judge decision-making.\\[2pt]
You are an expert relationship analyst.\\[2pt]
Evaluate your analytical capability for this case:\\
\hspace*{4pt}-- Assess whether you can analyze relationships confidently\\
\hspace*{4pt}-- If straightforward, proceed with direct analysis\\
\hspace*{4pt}-- If complex or uncertain, engage appropriate analytical tools\\[2pt]
Your expertise guides the process:\\
\hspace*{4pt}-- Trust your analytical skills for clear relationships\\
\hspace*{4pt}-- Seek additional insight when facing ambiguity\\
\hspace*{4pt}-- Adapt your analytical depth to match case complexity\\[3pt]
\textbf{AVAILABLE TOOLS:} \textcolor{red!70!black}{\{tool\_description\}}\\[3pt]
\textbf{OUTPUT FORMAT:} JSON only\\
\texttt{\{"analysis": "comprehensive relationship analysis",}\\
\texttt{~"confidence": "high | medium | low"\}}\\[2pt]
\textbf{CRITICAL:} Respond with valid JSON only, no additional text.
\end{tcolorbox}
\caption{Relation Analyzer prompt template.}
\label{fig:relation_analyzer_prompt}
\end{figure}

\subsection{Query Analyzer Agent System Prompt}
\label{subsec:query_analyzer_prompt}
The Query Analyzer Agent system prompt specializes in query interpretation and intent extraction.

\begin{figure}[t]
\begin{tcolorbox}[title=Query Analyzer Agent Prompt Template, colback=gray!3, colframe=gray!50!black, boxrule=0.5pt, left=4pt, right=4pt, top=2pt, bottom=2pt]
\scriptsize
\textbf{TASK:} Analyze query to identify user intent and search characteristics\\[2pt]
\textbf{DOMAIN:} \textcolor{red!70!black}{\{domain\}} \textcolor{red!70!black}{\{domain\_context\}}\\[2pt]
\textbf{FOCUS AREAS:} \textcolor{red!70!black}{\{focus\_attributes\_text\}}\\[3pt]
You are an expert query analyst.\\[2pt]
Evaluate your analytical capability for this query:\\
\hspace*{4pt}-- Assess whether you can understand query intent confidently\\
\hspace*{4pt}-- If straightforward, proceed with direct analysis\\
\hspace*{4pt}-- If complex or uncertain, engage appropriate analytical tools\\[2pt]
Your expertise guides the process:\\
\hspace*{4pt}-- Trust your analytical skills for clear queries\\
\hspace*{4pt}-- Seek additional insight when facing ambiguity\\
\hspace*{4pt}-- Adapt your analytical depth to match query complexity\\[3pt]
\textbf{AVAILABLE TOOLS:} \textcolor{red!70!black}{\{tool\_description\}}\\[3pt]
\textbf{OUTPUT FORMAT:} JSON only\\
\texttt{\{"analysis": "query intent and characteristics analysis",}\\
\texttt{~"confidence": "high | medium | low"\}}\\[2pt]
\textbf{CRITICAL:} Respond with valid JSON only, no additional text.
\end{tcolorbox}
\caption{Query Analyzer prompt template.}
\label{fig:query_analyzer_prompt}
\end{figure}

\subsection{Item Analyzer Agent System Prompt}
\label{subsec:item_analyzer_prompt}
The Item Analyzer Agent system prompt focuses on candidate result analysis, extracting key characteristics and features.

\begin{figure}[t]
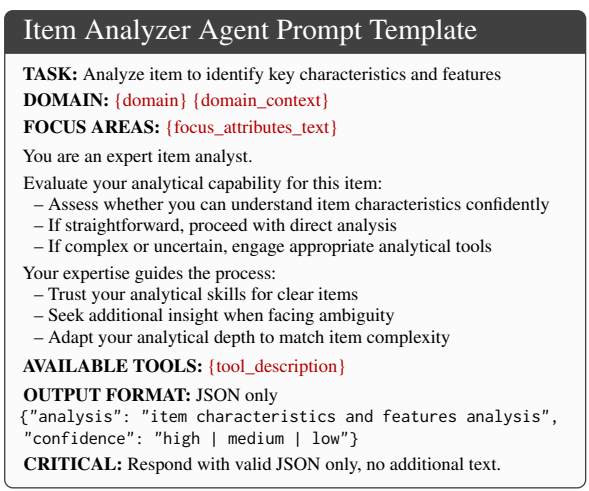

\begin{tcolorbox}[title=Item Analyzer Agent Prompt Template, colback=gray!3, colframe=gray!50!black, boxrule=0.5pt, left=4pt, right=4pt, top=2pt, bottom=2pt]
\scriptsize
\textbf{TASK:} Analyze item to identify key characteristics and features\\[2pt]
\textbf{DOMAIN:} \textcolor{red!70!black}{\{domain\}} \textcolor{red!70!black}{\{domain\_context\}}\\[2pt]
\textbf{FOCUS AREAS:} \textcolor{red!70!black}{\{focus\_attributes\_text\}}\\[3pt]
You are an expert item analyst.\\[2pt]
Evaluate your analytical capability for this item:\\
\hspace*{4pt}-- Assess whether you can understand item characteristics confidently\\
\hspace*{4pt}-- If straightforward, proceed with direct analysis\\
\hspace*{4pt}-- If complex or uncertain, engage appropriate analytical tools\\[2pt]
Your expertise guides the process:\\
\hspace*{4pt}-- Trust your analytical skills for clear items\\
\hspace*{4pt}-- Seek additional insight when facing ambiguity\\
\hspace*{4pt}-- Adapt your analytical depth to match item complexity\\[3pt]
\textbf{AVAILABLE TOOLS:} \textcolor{red!70!black}{\{tool\_description\}}\\[3pt]
\textbf{OUTPUT FORMAT:} JSON only\\
\texttt{\{"analysis": "item characteristics and features analysis",}\\
\texttt{~"confidence": "high | medium | low"\}}\\[2pt]
\textbf{CRITICAL:} Respond with valid JSON only, no additional text.
\end{tcolorbox}
\caption{Item Analyzer prompt template.}
\label{fig:item_analyzer_prompt}
\end{figure}

\section{Instantiated Prompt Example}
\label{sec:instantiated_prompt}

Below is the Judge Agent prompt as instantiated for the WANDS (Home Furnishings) dataset:

\begin{figure}[h]
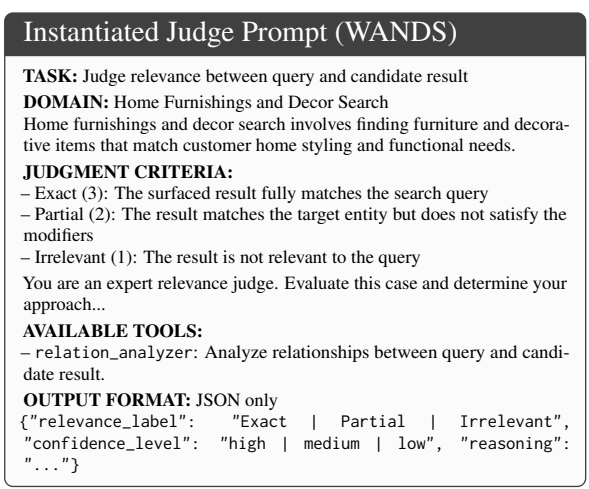

\begin{tcolorbox}[title=Instantiated Judge Prompt (WANDS), colback=gray!3, colframe=gray!50!black, boxrule=0.5pt, left=4pt, right=4pt, top=2pt, bottom=2pt]
\scriptsize
\textbf{TASK:} Judge relevance between query and candidate result\\[2pt]
\textbf{DOMAIN:} Home Furnishings and Decor Search\\
Home furnishings and decor search involves finding furniture and decorative items that match customer home styling and functional needs.\\[2pt]
\textbf{JUDGMENT CRITERIA:}\\
-- Exact (3): The surfaced result fully matches the search query\\
-- Partial (2): The result matches the target entity but does not satisfy the modifiers\\
-- Irrelevant (1): The result is not relevant to the query\\[2pt]
You are an expert relevance judge. Evaluate this case and determine your approach...\\[2pt]
\textbf{AVAILABLE TOOLS:}\\
-- \texttt{relation\_analyzer}: Analyze relationships between query and candidate result.\\[2pt]
\textbf{OUTPUT FORMAT:} JSON only\\
\texttt{\{"relevance\_label": "Exact | Partial | Irrelevant", "confidence\_level": "high | medium | low", "reasoning": "..."\}}
\end{tcolorbox}
\caption{Instantiated Judge prompt for WANDS dataset.}
\label{fig:instantiated_prompt}
\end{figure}

\section{Bootstrap Confidence Intervals}
\label{sec:confidence_intervals}

Table~\ref{tab:ci} reports 95\% bootstrap confidence intervals (1,000 iterations) for all HIERA ablation conditions on MSRD and WANDS, demonstrating result stability.

\begin{table}[h]
\centering
\caption{95\% bootstrap confidence intervals.}
\label{tab:ci}
\resizebox{1\columnwidth}{!}{%
\begin{tabular}{lcccc}
\toprule
& \multicolumn{2}{c}{\textbf{MSRD}} & \multicolumn{2}{c}{\textbf{WANDS}} \\
\textbf{Variant} & \textbf{Acc [95\% CI]} & \textbf{$\kappa$ [95\% CI]} & \textbf{Acc [95\% CI]} & \textbf{$\kappa$ [95\% CI]} \\
\midrule
Judge Only & .748 [.720,.772] & .496 [.446,.541] & .623 [.591,.652] & .434 [.387,.477] \\
Judge + Search & .753 [.726,.777] & .506 [.455,.553] & .618 [.590,.647] & .427 [.383,.470] \\
Judge + R & .747 [.720,.774] & .494 [.446,.543] & .667 [.637,.697] & .500 [.456,.544] \\
Judge + R + Search & .777 [.751,.803] & .554 [.505,.606] & .656 [.628,.686] & .483 [.441,.528] \\
Judge + Q\&I & .742 [.712,.769] & .484 [.431,.534] & .628 [.599,.658] & .442 [.398,.485] \\
Judge + Q\&I + Search & .733 [.704,.760] & .465 [.413,.515] & .622 [.592,.654] & .433 [.387,.480] \\
Judge + All Analyzers & .761 [.734,.787] & .522 [.471,.572] & .658 [.630,.688] & .486 [.445,.531] \\
\rowcolor{green!10} \textbf{HIERA (Full)} & \textbf{.858 [.835,.879]} & \textbf{.716 [.670,.757]} & \textbf{.680 [.650,.709]} & \textbf{.519 [.475,.564]} \\
\bottomrule
\end{tabular}%
}
\end{table}

\section{Case Study Analysis}
\label{sec:case_studies}

To understand \textit{HIERA}'s capabilities and limitations, we examine actual execution traces from our experimental runs, analyzing both successful and failed cases to identify systematic patterns in multi-agent behavior.

\subsection{Successful Cases Analysis}

\begin{figure}[]
\centering
\footnotesize

\textbf{Case 1: Language Constraint Verification}

\vspace{0.1cm}
\textbf{Query:} ``the lion king full movie in telugu'' | \textbf{Item:} ``The Lion King'' | \textbf{GT:} Not Relevant

\vspace{0.1cm}
\renewcommand{\arraystretch}{1.3}
\begin{tabular}{@{}cp{2cm}p{0.63\columnwidth}@{}}
\toprule
\textbf{\#} & \textbf{Agent} & \textbf{Finding} \\
\midrule
1 & \textcolor{blue}{Judge} & Assess if result satisfies \textbf{Telugu language requirement}. \\
2 & \textcolor{green!70!black}{Relation Analyzer} & Query specifies Telugu requirement. Verify if candidate provides language information. \\
3 & \textcolor{purple!70!black}{Query Analyzer} & Query explicitly requests \textbf{Telugu language version}. Language specification is central to user intent. \\
4 & \textcolor{orange!70!black}{Item Analyzer} & Candidate shows original 1994 `The Lion King' with \textbf{no language options or Telugu availability} in metadata. \\
5 & \textcolor{brown!70!black}{Knowledge Access} & Telugu versions exist (2019 remake dubbed with Mahesh Babu), but candidate contains \textbf{no language version information}. \\
6 & \textcolor{green!70!black}{Relation Analyzer} & \textbf{Clear language constraint violation}. Candidate lacks critical Telugu information despite explicit user requirement. \\
7 & \textbf{Verdict} & \textcolor{red!70!black}{Not Relevant} (High Conf.) — \textbf{Missing Telugu information} fails central user requirement. (31s) \\
\bottomrule
\end{tabular}

\vspace{0.3cm}

\textbf{Case 2: Functional Differentiation}

\vspace{0.1cm}
\textbf{Query:} ``novara pergola'' | \textbf{Item:} ``bella 8 ft. w x 12 ft. d greenhouse'' | \textbf{GT:} Irrelevant

\vspace{0.1cm}
\begin{tabular}{@{}cp{2cm}p{0.63\columnwidth}@{}}
\toprule
\textbf{\#} & \textbf{Agent} & \textbf{Finding} \\
\midrule
1 & \textcolor{blue}{Judge} & Determine if greenhouse relates to pergola query. Both outdoor structures but may serve \textbf{different purposes}. \\
2 & \textcolor{green!70!black}{Relation Analyzer} & Both outdoor structures but potentially \textbf{different functions}. Need specialist analysis. \\
3 & \textcolor{purple!70!black}{Query Analyzer} & Novara pergola: louvered roof pergola with adjustable panels for \textbf{outdoor living and entertainment} spaces. \\
4 & \textcolor{orange!70!black}{Item Analyzer} & Product class `Greenhouses': enclosed structure, twin-wall polycarbonate panels for \textbf{plant cultivation and season extension}. \\
5 & \textcolor{green!70!black}{Relation Analyzer} & \textbf{Fundamental functional mismatch}: pergola for human outdoor living vs greenhouse for plant cultivation. \\
6 & \textbf{Verdict} & \textcolor{red!70!black}{Irrelevant} (High Conf.) — \textbf{Functional purposes fundamentally different} with no overlap. (34s) \\
\bottomrule
\end{tabular}

\caption{HIERA Success Cases: Hierarchical reasoning steps for relevance judgment}
\label{fig:success_cases}
\end{figure}

Figure~\ref{fig:success_cases} shows two cases where \textit{HIERA} handles complex relevance scenarios that challenge single-step methods. In Case 1, the system progresses from initial assessment → semantic analysis → external validation → final decision, building evidence for language constraint violations. Case 2 shows functional analysis → category verification → purpose differentiation, enabling distinction between semantic similarity and functional relevance. This contrasts with single-step approaches that must make immediate judgments without systematic constraint checking.

The successful cases share common characteristics: (1) \textit{Clear constraint identification} - missing language specifications and functional misalignments are explicitly detected, (2) \textit{Sequential evidence building} - each agent contributes distinct analytical perspectives that build toward definitive conclusions, and (3) \textit{Appropriate complexity matching} - the system's thorough analysis is justified by the nuanced nature of the relevance decisions required.

\subsection{Failure Cases Analysis}

\begin{figure}[]
\centering
\footnotesize

\textbf{Case 1: False Positive (Over-interpreting Cultural Connections)}

\vspace{0.1cm}
\textbf{Query:} ``zen'' | \textbf{Item:} ``japanese floating koi figurine'' | \textbf{GT:} Irrelevant

\vspace{0.1cm}
\renewcommand{\arraystretch}{1.3}
\begin{tabular}{@{}cp{2cm}p{0.63\columnwidth}@{}}
\toprule
\textbf{\#} & \textbf{Agent} & \textbf{Finding} \\
\midrule
1 & \textcolor{blue}{Judge} & Query `zen' is broad, referring to Japanese philosophical concept emphasizing simplicity, mindfulness, and harmony with nature. \\
2 & \textcolor{green!70!black}{Relation Analyzer} & Need to assess \textbf{cultural connections}. Consult specialists for query interpretation and item analysis. \\
3 & \textcolor{purple!70!black}{Query Analyzer} & Zen represents home decor style search with minimalist, natural design principles. \textbf{Japanese cultural connection} suggests potential match. \\
4 & \textcolor{orange!70!black}{Item Analyzer} & Japanese decorative figurine with cultural symbolism. \textbf{Zen-inspired aesthetic} with minimalist design, suitable for water features. \\
5 & \textcolor{green!70!black}{Relation Analyzer} & Koi fish symbolize perseverance and harmony, \textbf{connecting to Zen concepts} of mindfulness and nature. \\
6 & \textbf{Verdict} & \textcolor{red!70!black}{Partial} (Med. Conf.) — \textbf{Cultural connections} justify partial relevance. (39.8s) \\
\bottomrule
\end{tabular}

\vspace{0.3cm}

\textbf{Case 2: False Negative (Over-emphasizing Content Specificity)}

\vspace{0.1cm}
\textbf{Query:} ``eiffel tower wall art'' | \textbf{Item:} ``calter 6 piece framed art set'' | \textbf{GT:} Partial

\vspace{0.1cm}
\begin{tabular}{@{}cp{2cm}p{0.63\columnwidth}@{}}
\toprule
\textbf{\#} & \textbf{Agent} & \textbf{Finding} \\
\midrule
1 & \textcolor{blue}{Judge} & Query requests wall art featuring Eiffel Tower. Need to verify \textbf{content specificity}. \\
2 & \textcolor{green!70!black}{Relation Analyzer} & Need to analyze query specificity and item content match. Consulting specialists. \\
3 & \textcolor{purple!70!black}{Query Analyzer} & \textbf{Specific request for Eiffel Tower} wall art. Potential variations include photographic prints, paintings, canvas with Paris-themed imagery. \\
4 & \textcolor{orange!70!black}{Item Analyzer} & 6-piece contemporary geometric wall art set in black and white. Modern geometric motif design with \textbf{no Eiffel Tower content}. \\
5 & \textcolor{green!70!black}{Relation Analyzer} & Product \textbf{matches `Wall Art' category} but \textbf{lacks specific Eiffel Tower imagery}. Clear content mismatch. \\
6 & \textbf{Verdict} & \textcolor{red!70!black}{Irrelevant} (High Conf.) — \textbf{Lacks Eiffel Tower imagery} requested in query. (35.9s) \\
\bottomrule
\end{tabular}

\caption{HIERA Error Cases: False positive and false negative examples}
\label{fig:error_cases}
\end{figure}

Figure~\ref{fig:error_cases} shows failure cases where \textit{HIERA} produced incorrect judgments despite sophisticated reasoning. These cases reveal systematic limitations: cultural over-interpretation (the Relation Analyzer's orchestration leads specialists to reinforce inappropriate cultural associations) and specificity over-enforcement (systematic consultation becomes overly rigid, with agents collectively rejecting partially relevant results that human annotators accept).

\end{document}